\documentclass[epj]{svjour}
\usepackage{latexsym}
\usepackage{graphicx}
\usepackage[usenames,dvipsnames]{xcolor} 
\usepackage{amsmath,amssymb}

\begin{document}
\title{Challenging event horizons with spin (3/2) fields}
\author{Koray D\"{u}zta\c{s}
}                     
%

\institute{Faculty of Engineering and Natural Sciences, \.{I}stanbul Okan University, 34959 Tuzla \.{I}stanbul, Turkey}

\date{Received: date / Revised version: date}
%
\abstract{We attempt to destroy the event horizons of Kerr black holes by perturbing them with massless spin (3/2) fields. We carry out a detailed analysis by incorporating the explicit form of the absorption probabilities and backreaction effects due to the self energy of the test fields.  For extremal and nearly extremal black holes, backreaction effects dominate for perturbations with large magnitudes. However, small perturbations can destroy the event horizons of extremal black holes and drive nearly extremal black holes closer to extremality.  Eventually, nearly extremal black holes reach a certain stage where they can be continuously driven to extremality and beyond. Both the cosmic censorship conjecture and the third law of black hole dynamics can be violated by spin (3/2) fields. This directly follows from the fact that fermionic fields do not satisfy the null energy condition. Therefore this result does not contradict with the fact that cosmic censorship and the laws of black hole mechanics remain valid for perturbations satisfying the null energy condition.
%
} 
\maketitle
\section{Introduction}
In a geodesically complete space-time, null geodesics can be extended to arbitrarily large values of the affine parameter both into the future and into the past. However, gravitation can bend nearby geodesics forcing them to eventually develop a focal point. In his seminal work, Penrose showed that this is a generic feature of classical general relativity \cite{pensing}. Null geodesics become future-incomplete after the development of a trapped surface. Subsequently, Hawking proved that time-like geodesics are past-incomplete in an expanding universe \cite{hawksing}. Respectively, the incompleteness of the null and time-like geodesics are interpreted as the existence of singularities inside the black holes, and the Big Bang singularity. 

If the singularities that generically ensue as a result of gravitational collapse are shrouded behind event horizons of black holes, they cannot have a causal contact with the observers at the asymptotically flat infinity. The fact that the event horizons act as a one-way membrane for causal curves, disables this causal contact. The smooth causal structure of the space-time can be preserved excluding the black region. On the contrary, if the singularity is \emph{naked} there exist time-like curves into the past that terminate on the singularity. One tends to think that there must be a natural mechanism to prevent the formation of naked singularities. In that respect, Penrose proposed the cosmic censorship conjecture which --in its weak form-- states that the singularities that ensue as a result of gravitational collapse are always hidden behind the event horizons of black holes \cite{ccc}. The deterministic nature of classical general relativity relies on the validity of this conjecture which still lacks a rigorous proof. The subsequent developments in black hole physics such as the laws of black hole thermodynamics, Hawking radiation, the holographic principle also pre-assume the existence of the event horizon. 

Not only the formation of naked singularities in gravitational collapse, but also the possibility of the evolution of black holes to naked singularities would be problematic for black hole physics. One can check the possibility to destroy the event horizon and expose the singularity to outside observers, by constructing thought experiments. In these thought experiments one starts with a black hole and perturbs it with test particles or fields to check the possibility to drive the black hole beyond the extremal point; i.e. destroy the event horizon. For example the Kerr metric represents a black hole surrounded by an event horizon if the mass $(M)$ and angular momentum $(a)$ parameters satisfy the inequality $M^2 \geq a^2$. If one can increase the angular momentum parameter to exceed the mass parameter by sending in test particles or fields from infinity, the event horizon ceases to exist. In these thought experiments one assumes that the contribution of the test particles or fields to the space-time parameters  are small so that  the structure of the metric does not change. The space-time geometry is represented by the Kerr metric (or the relevant initial metric) before and after the interaction. The first thought experiment in this form was constructed by Wald \cite{wald74}. It appears that the particles and fields carrying sufficiently large charge and angular momentum to lead to the destruction of the event horizon are not absorbed by  black holes. Test particles and bosonic fields cannot destroy  event horizons considering the backreaction effects, if necessary \cite{hu,js,backhu,backjs,f1,gao,siahaan,magne,yuwen,higher,v1,he,wang,jamil,shay3,shay4,zeng,semiz,q1,q2,q3,q4,q5,q6,q7,overspin,emccc,natario,duztas2,mode,taubnut,kerrsen,hong,yang,kerrmog,bai,khoda,ong}. 
The argument also applies to the asymptotically de-Sitter and anti-de Sitter cases \cite{btz,gwak3,chen,ongyao,mtz,he2,dilat,yin,btz1,yang1,shay5,gwak4,shay6,corelli,sia2}.

For bosonic test fields there exists a lower limit for the frequency to allow the absorption of the field by the black hole. The fields with a frequency lower than the limiting value are reflected back to infinity with a larger amplitude in a process known as superradiance. In a static and axi-symmetric space-time the wave equations for test fields are separable, which comprises spin (3/2) fields in Kerr background \cite{guven}. The energy ($\delta E$) and the angular momentum ($\delta J$) carried by the field are related by:
\begin{equation}
\delta J= \frac{m}{\omega} \delta E
\label{deltaj}
\end{equation}
where $m$ is the azimuthal wave number, and $\omega$ is the frequency of the test field.  As the frequency of the test field decreases its contribution to the angular momentum parameter of the black hole increases relative to its energy which contributes to the mass parameter. ( $\delta J$ is inversely proportional to $\omega$.) Thus, the test fields with low frequencies can lead to overspinning of the black holes into naked singularities. Fortunately, the existence of the superradiance limit forbids the absorption of test fields with low frequencies, in favour of cosmic censorship. However a corresponding lower limit does not exist for fermionic fields. The absorption of test fields with low frequency is allowed, which leads to drastic results regarding the validity of cosmic censorship \cite{duztas,toth,generic}. As $\omega$ approaches zero, the difference $(M^2-a^2)$ --which should be non-negative for the existence of the event horizon-- diverges to minus infinity. This drastic divergence problem is solved by incorporating the explicit form of the absorption probabilities into the problem, in an original approach which we have suggested in \cite{absorp} and applied to spin (1/2) fields in \cite{spinhalf}. In this approach we consider the fact that the interaction of test fields is a scattering problem. The test field is partially absorbed by the black hole and partially reflected back to infinity. The fraction of the test field absorbed by the black hole is given by the absorption probability. If the initial mass parameter of the black hole is $M$, the final mass parameter is given by
\begin{equation}
M_{\rm{fin}} = M+\Gamma (\delta M )
\end{equation}
where $\delta M$ is the energy of the field, and $\Gamma$ is the absorption probability. In the conventional approach one assumes $(\Gamma \sim 1)$ provided that it is positive. A more subtle analysis of the problem requires the incorporation of the explicit form of the absorption probability. For example, the absorption probability of spin (1/2) fields depends on $\omega^2$. As $\omega$ approaches zero, the absorption probability also approaches zero, which indicates that the test field is entirely reflected back to infinity leaving the space-time parameters unaltered. If one assumes $(\Gamma \sim 1)$ for these modes, the angular momentum parameter of the black hole increases to arbitrarily large values and the quantity $(M^2-a^2)$ diverges to minus infinity. In \cite{generic} we showed that this is indeed the case for spin (1/2) fields. The quantity $(M^2-a^2)$ becomes zero near $\omega \sim 0.25 (1/M)$, and it diverges to minus infinity as $\omega$ decreases. However, when one incorporates the absorption probabilities and the backreaction effects into the analysis the overspinning problem can be fixed for fields with a relatively large amplitude and becomes rather non-generic for fields with small amplitudes \cite{spinhalf}.

In this work we consider the interaction of massless spin (3/2) fields with Kerr black holes to test the validity of the cosmic censorship conjecture. If one adapts the conventional approach assuming $\Gamma \sim 1$, the analysis is almost identical with the case of spin (1/2) fields except substituting $m=(3/2)$ for the azimuthal wave number.   This time, the quantity $(M^2-a^2)$ becomes zero near $\omega \sim 0.75 (1/M)$, and sharply diverges to minus infinity. (See Equation (12) in \cite{spinhalf}) Here we carry out a more subtle analysis by incorporating the explicit form of the absorption probabilities. For that purpose we use the absorption probability derived by Page \cite{page}. First we test the validity of the cosmic censorship conjecture for extremal black holes by incorporating the absorption probabilities and backreaction effects. For nearly extremal black holes we evaluate the validity of the cosmic censorship conjecture and the third law of black hole dynamics, simultaneously. Finally, we summarise and interpret our results.

\section{Spin (3/2) fields and cosmic censorship}
\label{ext}
We start our analysis by deriving the expression for the absorption probability of spin (3/2) fields based on the general formula derived by Page \cite{page}. Page derived the absorption probabilities $\Gamma_{s\omega lm}$ for test fields of spin $s$, frequency $\omega$, spherodial harmonic $l$, and azimuthal wave number $m$. For fermionic fields the general expression for the absorption probability is
\begin{equation}
\Gamma=\left[ \frac{(l-s)!(l+s)!}{(2l)!(2l+1)!!}\right]^2 \left(\frac{A\kappa}{2\pi}\omega \right)^{2l+1} \prod_{n=1}^{(l+1/2)}\left[1+\left(\frac{\omega - m\Omega}{n\kappa -\frac{1}{2}\kappa}\right)^2 \right]
\label{pageprob}
\end{equation}
(see equation (14) in \cite{page}). Here, $\Omega$ is the angular velocity of the horizon, $\kappa$ is the surface gravity, $A$  is the area of the event horizon. Explicitly,
\[ \Omega=\frac{a}{r_+^2+a^2} 
;\kappa=\frac{r_+-r_-}{2(r_+^2+a^2)} ;
 A=4\pi (r_+^2+a^2) \]
We are going to evaluate each term in (\ref{pageprob}) separately, so that the reader can easily verify the calculation. The dominant contribution comes from the $l=s$ modes. By substituting $l=s=(3/2)$ one finds
\[ 
\left[ \frac{(l-s)!(l+s)!}{(2l)!(2l+1)!!}\right]^2 =\frac{1}{64}
\]
Substituting  $l=(3/2)$,
\[
\Gamma= \frac{1}{64}\left(\frac{A\kappa}{2\pi}\omega \right)^4 \left[1+\frac{4(\omega - m\Omega)^2 }{\kappa^2} \right] \left[1+\frac{4(\omega - m\Omega)^2 }{9\kappa^2} \right]
\]
where we have substituted $n=1$ and $n=2$ in the product term. This can be written as
\begin{equation}
\Gamma= \frac{A^4 \omega^4}{64(2\pi)^4}\left[\kappa^4 +\frac{40\kappa^2(\omega - m\Omega)^2 }{9}+ \frac{16(\omega - m\Omega)^4}{9}\right]
\label{prob1}
\end{equation}
Keeping the lowest order terms in $\omega$, the absorption probability can be further simplified to give
\begin{equation}
\Gamma  = \frac{1}{4} ( M^4 +8M^2a^2 ) \omega^4 
\label{probpage} 
\end{equation}
However this result pre-assumes $\omega \ll m\Omega$. For modes with $\omega \sim m\Omega$, one should refer to the absorption probability (\ref{prob1}).  The absorption probability for spin (3/2) fields is positive definite for every mode. It rapidly converges to zero for low frequency modes which implies that low frequency modes are almost entirely reflected back to infinity.

\subsection{Destroying extremal black holes}
We first evaluate the validity of the cosmic censorship conjecture for extremal Kerr black holes interacting with spin (3/2) test fields. Initially the mass and angular momentum parameters of the space-time satisfy
\begin{equation}
M^2-J=0
\label{paramnext}
\end{equation}
We envisage a test spin (3/2) field that is incident on the black hole from infinity. The field carries energy $\delta M$ and angular momentum $\delta J=(m/\omega) \delta M$, where $m=(3/2)$. The field is partially absorbed by the black hole and partially reflected back to infinity. In the final case, the fraction absorbed by the black hole modifies the mass and angular momentum parameters. This fraction is determined by the absorption probability of the field. The final parameters take the form
\begin{eqnarray}
&&M_{\rm{fin}}=M +\Gamma \delta M \nonumber \\
&&J_{\rm{fin}}=J +\frac{m}{\omega}\Gamma \delta M
\label{extparam}
\end{eqnarray}
One observes that ignoring the absorption probability --which is equivalent to setting $\Gamma \sim 1$ -- would lead the final value of the angular momentum parameter to increase without bound as $\omega$ approaches zero. However, by incorporating the absorption probability, one finds that the mass angular momentum parameters attain their initial values as $\omega$ approaches zero. The physical interpretation is clear. The absorption probability also approaches zero faster than the frequency $\omega$ itself. In that case, the field is entirely reflected back to infinity, so the space-time parameters remain invariant. Though the  the drastic divergence problem of the angular momentum parameter does not arise when one incorporates the absorption probabilities, the possibility of overspinning black holes may persist. 

The absorption probability for extremal black holes takes the form:
\begin{equation}
\Gamma=\frac{64}{9}M^8 \omega^4 (\omega - \frac{3}{4M})^4
\label{probext}
\end{equation}
which can be derived by substituting, $\kappa=0$, $A=8\pi M^2$, $\Omega=(1)/(2M)$, and $m=3/2$ in (\ref{prob1}).
We are going to perturb an extremal black hole with a mode which satisfies $\omega < m\Omega$ since these modes contribute to the angular momentum parameter more than the mass parameter. Among these modes we are going to choose the one with the highest absorption probability. For the relevant modes, the absorption probability (\ref{probext}) attains its maximum value at
\begin{equation}
\omega=\frac{3}{8M}
\end{equation}
The maximum value of the absorption probability can be calculated:
\begin{equation}
\Gamma\left( \frac{3}{8M}\right)=0.0028
\label{probnum}
\end{equation}
For the energy of the test field we let $\delta M=M\eta$, where $\eta \ll 1$. This ensures that the contribution of the test field to the space-time parameters is of the first order and the test field approximation is justified. We define the quantity
\begin{equation}
\Delta_{\rm{fin}}=M_{\rm{fin}}^2-J_{\rm{fin}}
\end{equation} 
If $\Delta_{\rm{fin}}$ becomes negative at the end of the interaction, we conclude that the black hole has been overspun  into a naked singularity. Substituting  $\omega=(3)/(8M)$ in (\ref{extparam}), the final value of the angular momentum parameter takes the form
\begin{equation}
J_{\rm{fin}}=J+4M^2 \Gamma \eta
\end{equation}
Making the necessary substitutions we can express $\Delta_{\rm{fin}}$ as  
\begin{eqnarray}
\Delta_{\rm{fin}} &=& M^2 + M^2\Gamma^2 \eta^2 +2M^2 \Gamma \eta -J - 4M^2 \Gamma \eta \nonumber \\
&=& M^2\Gamma^2 \eta^2 -2M^2 \Gamma \eta
\label{deltafinext}
\end{eqnarray}
For small $\eta$, the result (\ref{deltafinext}) is negative definite, which implies that extremal  black holes can be overspun into a naked singularities as a result of their interactions with spin (3/2) test fields. On the other hand a second order calculation is incomplete without incorporating the backreaction effects. We proceed by calculating the backreaction effects.

\subsection{Backreaction effects}
In his seminal paper Will argued that a test field  incident on a black hole  will  induce an increase on the angular velocity of the event horizon before the field is absorbed by the black hole \cite{will}. To avoid any confusion, we note that the increase in the angular velocity of the horizon is not due to an increase in the angular momentum of the black hole. The mass and the angular momentum parameters of the black hole do not change before the absorption of the field. Only, the angular velocity of the horizon increases. The induced increase in the angular velocity was estimated by Will 
\begin{equation}
\Delta \Omega = \frac{\delta J}{4M^3}
\end{equation}
where $\delta J$ is the angular momentum of the test field   This induced increase in the angular velocity leads to a first order correction in the self energy of the field.
\begin{equation}
E^{(1)}_{\rm{self}}=(\Delta \Omega)(\delta J)=\frac{(\delta J)^2}{4M^3}
\label{self}
\end{equation}
The self energy of the test field contributes to the mass parameter of the black hole and modifies $M_{\rm{fin}}$ and $\Delta_{\rm{fin}}$. Both the increase in the angular velocity of the horizon and the self energy of the test field act as backreaction effects in the scattering problem. The increase in the angular velocity slightly modifies the absorption probability  as we substitute $\Omega$ by
\[ \Omega'= \Omega + \delta \Omega
\]
For the mode with the highest absorption probability ($\omega=3/(8M)$, $\delta J=4M^2 \eta$) one finds that
\[ \Delta \Omega = \eta \frac{1}{M} \]
For example, if we let $\eta=0.001$, the angular velocity increases from $\Omega=(1)/(2M)$ to $\Omega'=(0.501)/M$. This does not change the absorption probability (\ref{probnum}) to four digits after the decimal.
Though its effect is small, the increase in the angular velocity has a unique feature as a backreaction effect. As it slightly increases the absorption probability, it does not prevent but reinforces the probability of overspinning. This is also the case for spin (1/2) fields \cite{spinhalf}.

We are going to calculate the self energy for the mode with the highest absorption probability.
\begin{equation}
E^{(1)}_{\rm{self}}=\frac{(\delta J)^2}{4M^3}=4M \eta^2
\label{self33}
\end{equation}
As expected the self energy  contributes to the mass parameter to second order in $\eta$. The final value of the mass parameter is modified as
\begin{equation}
M_{\rm{fin}}=M + M\Gamma \eta + 4M\eta^2
\end{equation}
Now we re-calculate $\Delta_{\rm{fin}}$  using the modified value of $M_{\rm{fin}}$ after incorporating the backreaction effects. 
\begin{equation}
\Delta_{\rm{fin}}=M^2 (\Gamma^2 \eta^2 -2  \Gamma \eta + 8 \eta^2) +O(\eta^3)
\label{deltafinback}
\end{equation}
where $\Gamma=0.0028$. (\ref{deltafinback}) implies that for large perturbations of the order $\eta \sim 0.001$, $\Delta_{\rm{fin}}$ can be positive after the interaction; i.e. the incorporation of the backreaction effects can fix the overspinning problem. However for smaller perturbations of the order $\eta \lesssim 0.0001$, $\Delta_{\rm{fin}}$ will still be negative after incorporating the backreaction effects. 

\section{Nearly extremal black holes and the third law}
\label{next}
The laws of black hole dynamics were proposed by Bardeen, Carter, and Hawking \cite{bch} who attempted to set up an analogy between black hole physics and the laws of thermodynamics. According to their statement of the third law, no procedure can reduce the surface gravity $\kappa$ to zero by a finite sequence of operations. They also argue that, if this was possible, the process could be presumably carried further to create a naked singularity. Later Israel established the formal statement of the third law which was accompanied by a proof \cite{Israel}. According to the formal statement, no continuous process in which the energy-momentum tensor of the accreted matter satisfies the weak energy condition can reduce the surface gravity to zero within a finite advanced time. Israel's proof is based on his gravitational confinement theorem which pre-assumes that trapped surfaces can be extended in an area-preserving fashion, provided that they do not encounter any singularities during their developments \cite{Israel2}. Previously we argued that these assumptions cannot be justified in a universe where the cosmic censorship hypothesis is violated. A concrete proof of the third law should not involve any direct or indirect assumption of cosmic censorship \cite{tjphys}.

Whether a concrete proof of the third law exists or not, its validity is justified for Kerr black holes perturbed by bosonic fields the energy-momentum tensor of which satisfy the weak energy condition. For test bodies \cite{js} and fields \cite{overspin} it was found that though extremal black holes cannot,  nearly extremal black holes can be overspun into naked singularities. However overspinning occurs by a discrete jump from a nearly extremal black hole to a naked singularity. If one attempts to drive the black hole continuously to extremality and beyond, one finds that the range of parameters for the perturbation that can be used to overspin the black hole pinches off before the extremality is reached,  which was also pointed out in \cite{dadhich}. The discrete jump from a nearly extremal black hole to  a naked singularity can be interpreted as an artefact of our definition of the problem. Moreover the overspinning of nearly extremal black holes is fixed by employing backreaction effects \cite{backjs,kerrmog}. Therefore one cannot decrease the surface gravity continuously to zero; i.e. the third law is valid if the perturbations satisfy the weak energy condition.

On the other hand, the energy momentum tensor for fermionic fields does not satisfy the weak energy condition. In this section we test the validity of the third law and the cosmic censorship conjecture for a nearly extremal black hole interacting with a spin (3/2) field. We start with a nearly extremal black hole which satisfies
\begin{equation}
M^2-J=M^2\epsilon_0^2
\end{equation}
where $\epsilon_0 \ll 1$ parametrizes the closeness to extremality. We perturb this black hole by a test field with frequency $\omega=3/(8M)$ and energy $\delta E=M\eta$ which was used in the previous section to overspin extremal black holes. We also incorporate the self-energy of the field. The calculations in the previous section imply that 
\begin{equation}
\Delta_{\rm{fin}}=M^2 (\Gamma^2 \eta^2 -2  \Gamma \eta + 8 \eta^2 +\epsilon_0^2)
\label{deltafinnext}
\end{equation}
For nearly extremal black holes $\kappa$ is small, therefore $\kappa^2$ and $\kappa^4$ terms do not considerably contribute to the absorption probability. The absorption probability $\Gamma$ is  equal to its value calculated for extremal black holes to four significant figures; i.e. $\Gamma=0.0028$. For large values of $\epsilon_0$ (such as $\epsilon_0 \sim 0.001$) the equation
\begin{equation}
\Gamma^2 \eta^2 -2  \Gamma \eta + 8 \eta^2 +\epsilon_0^2=0
\label{epsilon}
\end{equation}
has no real roots with $\Gamma=0.0028$, which implies that $\Delta_{\rm{fin}}$ cannot be zero or negative after the interaction. However we can perturb this nearly extremal black hole with test fields with $\eta \lesssim 0.0001$ for which 
\begin{equation}
(8+\Gamma^2) \eta^2 -2  \Gamma \eta <0 
\label{smallpert}
\end{equation}
Though $\Delta_{\rm{fin}}$ cannot be negative after this interaction its value will decrease. If
\[ M^2-J=M^2\epsilon_1^2 \]
after the interaction, $\epsilon_1$ will be less than $\epsilon_0$. After a finite number of successive perturbations, we will have $\epsilon_n \sim 0.0001$. The equation (\ref{epsilon}) can be analytically solved for $\eta$, if $\epsilon=0.0001$ and $\Gamma=0.0028$. It has two real roots $\eta_1=1.79 \times 10^{-6}$, and $\eta_2=6.9 \times 10^{-4}$. The value of $\Delta_{\rm{fin}}$ will be negative between the roots, and positive elsewhere. The value of $\eta$ can be adjusted such that $\Delta_{\rm{fin}}$ is positive, negative or zero. In other words, a nearly extremal black hole can be continuously driven  to extremality and beyond by spin (3/2) fields, in a finite number of steps. Therefore both the cosmic censorship conjecture and the third law of black hole dynamics can be violated by a judicious choice of frequency and magnitude for a spin (3/2) test field.

\section{ Comparison of bosonic and fermionic cases}
\label{compare}
At this stage it would also be appropriate to compare the results for fermionic and bosonic fields in order to resolve potential ambiguities. In an early work we had derived that --though extremal black holes cannot-- nearly extremal black holes can be overspun by bosonic fields \cite{overspin}. This overspinning is not generic; it can be fixed by the backreaction effects which were ignored in that work. In particular the backreaction effects due to Will, which are employed in this work can fix the overspinning problem for nearly extremal black holes. We have justified this in a recent paper on Kerr-MOG black holes \cite{kerrmog}. We showed that  backreaction effects fix the overspinning problems in the limit the MOG parameter $\alpha$ equals zero, which describes Kerr black holes. The range of frequencies that can overspin the black hole pinches off as extremality is approached, which implies that the third law is valid, in agreement with a previous work by Dadhich and Karayan \cite{dadhich}.  We also showed that cosmic censorship is valid in the case of spin-1; i.e. for electromagnetic test fields \cite{emccc}. The same result is expected for spin-2 fields, which we are going to evaluate in a future work. We can safely state that cosmic censorship conjecture remains valid for perturbations satisfying the null energy condition, in the context of classical general relativity. We have encountered some violations of the conjecture in alternative theories of gravity as in the cases of Kerr-MOG \cite{kerrmog} and charged dilaton \cite{dilat} black holes. However, these should be interpreted as evidence against the validity of the underlying theory leading, rather than the cosmic censorship itself. 

In black hole physics the correct statement that the cosmic censorship is valid for perturbations satisfying the null energy condition, is  usually attributed to a work by Sorce and Wald \cite{w2}. However in \cite{absorp} we scrutinized the method developed by Sorce and Wald, and explicitly demonstrated that it involves order of magnitude errors. Therefore this correct statement or any valid statement cannot be inferred by Sorce-Wald method. Despite the fact that the order of magnitude errors are manifest and we have pedagogically elucidated the subject in \cite{absorp}, Sorce-Wald method maintains its popularity in black hole physics.

Fermionic perturbations are problematic concerning the validity of the cosmic censorship conjecture and the laws of black hole dynamics. In our previous works we have derived that cosmic censorship can be violated by fermionic fields,  in the classical picture \cite{duztas,generic,spinhalf}. Independently, Toth also derived the same result for fermionic fields in \cite{toth}. There were also works involving particles some of which were fermions \cite{q1,q2,q3,q4,q5,q6}. We have meticulously analysed these works and resolved their ambiguities in \cite{q7}. 

The main difference between the fermionic and bosonic perturbations is that the energy momentum tensor of the former does not satisfy the weak/null energy condition. The null energy condition allows us to derive a lower limit for the energy of a particle or field to allow its absorption by the black hole. This lower limit can be generally expressed by Needham's condition \cite{needham}:
\begin{equation}
\delta M-\Omega_H \delta J-\Phi_H\delta Q \geq 0
\label{needham}
\end{equation}
where $\Omega_H=a/(r_+^2 +a^2)$, $\Phi_H =(Qr_+)/(r_+^2 +a^2)$, and $r_+$ is the horizon radius. The first derivation of this condition known to this author is by Needham in 1980. Sorce-Wald derive the same result without referring to previous works. Needham's condition determines the lower limit for various types of perturbations satisfying the null energy condition. For example for Reissner-Nordstrom black holes ($a=0$) perturbed by test particles with charge $\delta Q=q$, (\ref{needham}) gives:
\[
\delta M=E>\frac{qQ}{r_+}
\]
which is exactly the lower limit derived by Hubeny in \cite{hu}, who used different methods. In another well-known work by Jacobson and Sotiriou the lower limit for the energy of a test body to be absorbed by a Kerr black hole ($Q=0$) is calculated as \cite{js}
\[
E>\Omega_H \delta J
\]
which is implied by (\ref{needham}) with $Q=0$. Also for test fields in static and axi-symmetric spacetimes, carrying energy $\delta M=E$ and angular momentum $\delta J=(m/\omega) E$, (\ref{needham}) reduces to
\[
\omega \geq m\Omega
\]
which is precisely the superradiance condition which applies to bosonic test fields satisfying the null energy condition. 

The lower limit for the energy does not exist for fermionic fields due to the fact that they do not satisfy the null energy condition. Equivalently one can state this as the fermionic fields do not satisfy the Needham's condition, or they do not exhibit superradiant scattering, or the absorption probability is always positive. The absorption of low energy modes leads to drastic results regarding the validity of the cosmic censorship conjecture.  As the frequency (or the energy) of the fields decreases its contribution to the angular momentum and/or charge parameters increases relative to its contribution to the mass parameter. In the final case the angular momentum and/or charge parameters of the space-time settles to values larger than the mass parameter leading to the formation of a naked singularity. We have attempted to alleviate this tension by incorporating absorption probabilities. This renders the use of the modes with very low frequencies like $\omega <0.1 (1/M)$ inefficient. For these modes the absorption probability approaches zero, which means they are effectively reflected back to infinity. However, one can still find modes that can overspin extremal and near extremal black holes.  

Let us elucidate the difference between bosonic and fermionic fields with a numerical example. In sections (\ref{ext}) and (\ref{next}) we used test fields with frequencies $\omega=3/(8M)$ to overspin the black hole. If the stress energy tensor for fermionic fields did satisfy the null energy condition, these fields would obey Needham's condition (\ref{needham}). In that case the absorption of test fields with $\omega < m\Omega$ would not be allowed. For $m=3/2$ this would imply ( $\Omega=1/(2M)$ for an extremal black hole.)
\begin{equation}
\omega \geq (3/2) \Omega \geq \frac{3}{4M}
\end{equation}
The modes with $\omega < 3/(4M)$ would not be absorbed by the black hole. In a more subtle analysis the absorption probability  would be zero for $\omega = 3/(4M)$, and negative for smaller frequencies. In that case one would not be able to use the mode with $\omega=3/(8M)$ to overspin the black hole. Let us envisage how the thought experiment would be modified if spin 3/2 fields did satisfy the null energy condition. Assume we proceed with a test field with frequency slightly larger than $\omega = 3/(4M)$ and energy $\delta M=M\eta$. For that mode the contribution to the angular momentum parameter would be.
\begin{equation}
\delta J=(m/\omega) \delta M =2M^2 \eta
\end{equation}
where $m=3/2$. Starting with an extremal black hole with $M^2-J=0$ the final parameters of the space-time would satisfy:
\begin{equation}
M_{\rm{fin}}^2-J_{\rm{fin}}=(M+\Gamma M\eta)^2-J-2\Gamma M^2 \eta=\Gamma^2 M^2 \eta^2
\label{deltafinhypo}
\end{equation}
which implies that we have started with an extremal black hole and ended up with a nearly extremal black hole. Cosmic censorship remains valid without any need to employ backreaction effects. This also implies the validity of the third law since it is not possible to make the right hand side of (\ref{deltafinhypo}) negative in a small neighbourhood of the extremal state; i.e. one cannot continously drive the black hole to extremality.  For higher frequencies the right hand side of (\ref{deltafinhypo}) would be even larger, and the absorption of lower frequencies would not be allowed. The fact that the absorption of lower frequencies is allowed,  leads to problems for fermionic fields. The right hand side of the equation becomes negative as $\omega$ decreases; it can even diverge to $-\infty$ as $\omega \to 0$. This problem is solved by incorporating the explicit form of the absorption probabilities. However, we have not been able to preserve the validity of cosmic censorship. This may require a quantum treatment of the problem beyond the semi-classical approximations.

\section{Conclusions}
In this work, we have evaluated the possibility to overspin Kerr black holes into naked singularities by perturbing them with massless spin (3/2) fields. Previously we argued that the interaction of black holes with test fields is a scattering problem and the explicit form of the absorption probabilities should be included in a complete analysis \cite{absorp}. If one ignores the absorption probabilities --which corresponds to letting $\Gamma \sim 1$ whenever it is positive-- the overspinning of black holes due to fermionic fields is inevitable. Moreover, as the frequency of the test field approaches zero, the quantity $M^2-a^2$ diverges to minus infinity. The incorporation of the absorption probabilities solves this drastic divergence problem.

Supervening on our previous arguments, we carried out a detailed analysis by incorporating the explicit form of the absorption probabilities for spin (3/2) fields.  We used the general expression for absorption probabilities derived by Page \cite{page}. The absorption probability depends on the initial parameters of the black hole and the frequency of the test field. It does not explicitly depend on the magnitude of the perturbation. We started with an extremal Kerr black hole and perturbed it with a test field carrying energy $\delta E=M\eta$. For the frequency of the test field we chose the mode with the highest absorption probability. We also incorporated backreaction effects due to the self energy of the test field.  We found that, for perturbations of large magnitude ($\eta \gtrsim 0.001$), backreactions dominate and the event horizon cannot be destroyed. However, for smaller perturbations  overspinning is possible. An extremal black hole can be overspun into a naked singularity. 

Then we evaluated the validity of the cosmic censorship conjecture and the third law of black hole dynamics for nearly extremal black holes. We found that if the black hole is not sufficiently close to extremality, overspinning is not possible. However, small perturbations which satisfy (\ref{smallpert}), drive the black hole closer to extremality. After a finite number of steps the black hole becomes sufficiently close to extremality. At this stage it can be driven to extremality or overspun into a naked singularity by small perturbations. Therefore both the cosmic censorship conjecture and the third law of black hole dynamics can be violated by spin (3/2) test fields, by a judicious choice of frequency and magnitude.

Recently we have constructed a similar thought experiment with spin (1/2) fields. Spin (3/2) fields  contribute to the  angular momentum with a magnitude that is precisely three times larger than a spin (1/2) field with the same frequency. The modes become challenging around $\omega \lesssim m \Omega$, which implies $\omega \sim 0.75 (1/M)$ for spin (3/2) fields and $\omega \sim 0.25 (1/M)$ for spin (1/2) fields. One may expect the destruction of the event horizon to be more generic for spin (3/2) fields. However, the absorption probability decreases much  faster as $\omega$ decreases in the spin (3/2) case. Therefore the results for the two cases are analogous. Backreaction effects are dominant for large perturbations, and small perturbations can lead to overspinning.

We compared the results for bosonic and fermionic perturbations --satisfying and not satisfying the null energy condition--  in section \ref{compare}. The main difference is that there exists a lower limit for the energy of the perturbation to allow its absorption by the black hole if the null energy condition is satisfied. The absence of this lower limit leads to problems for fermionic perturbations. We mentioned that if spin 3/2 fields did satisfy the null energy condition the absorption of the modes $\omega \leq (3/2) \Omega$ would not be allowed. We also showed that the modes that would be absorbed by the black hole would not lead to the violation of cosmic censorship or the third law.

%
%

\end{document}